\documentclass[epj]{svjour}

\usepackage{graphicx}
\usepackage{dcolumn}
\usepackage{bm}
\usepackage{bbm}
\usepackage{psfrag}
\usepackage{multirow,amssymb,amsbsy,amsmath}
\usepackage{stmaryrd}
\usepackage[sort&compress,numbers]{natbib}

\newcommand{\I}{\textup{i}}
\newcommand{\E}{\textup{e}}
\newcommand{\Dk}{\hspace*{-6mm}\textup{d}}
\newcommand{\D}{\textup{d}}
\newcommand*{\Haver}[1]{\mathopen{\llbracket} #1 \mathclose{\rrbracket}}

\newcommand{\dd}{\text{d}}
\newcommand{\dod}[2]{\frac{\dd #1}{\dd #2}}

\newcommand{\Funktion}[2]{#1\kern-0.2em\left(#2\right)}

\newcommand{\trtxt}[2][]{\text{Tr}_{#1}\{#2\}}
\newcommand{\com}[2]{[#1,#2]}

\newcommand*{\bra}[1]{\mathopen{\langle}#1\mathclose{|}}
\newcommand*{\ket}[1]{\mathopen{|}#1\mathclose{\rangle}}
\newcommand*{\ketbra}[2]{\mathopen{|}#1\rangle\langle#2\mathclose{|}}
\newcommand*{\sprod}[2]{\mathopen{\langle}#1|#2\mathclose{\rangle}}

\newcommand{\refsec}[1]{Sect.~\ref{#1}}
\newcommand{\reffig}[1]{Fig.~\ref{#1}}

\newcommand{\proj}[2]{\hat{P}_{#1,#2}}

\begin{document}

%
%
\title{Finite quantum environments as thermostats: an analysis based on the Hilbert space average method}

\author{Jochen~Gemmer\inst{1} \and Mathias~Michel\inst{2}}
\institute{Physics Department, %
           Universit\"at Osnabr\"uck, %
           Barbarastr.\ 7, %
           D-49069 Osnabr\"uck, %
           Germany %
           \mail{jgemmer@uos.de} %
           \and %
           Institut f\"ur Theoretische Physik I, %
           Universit\"at Stuttgart, %
           Pfaffenwaldring 57, %
           D-70550 Stuttgart, %
           Germany}

\date{Received: \today / Revised version: date}

\authorrunning{J. Gemmer and M. Michel}
\titlerunning{Finite quantum environments as thermostats: an analysis based on HAM}

\abstract{
We consider discrete quantum systems coupled to finite environments which may possibly consist of only one particle in contrast to the standard baths which usually consist of continua of oscillators, spins, etc. 
We find that such finite environments may, nevertheless, act as thermostats, i.e., equilibrate the system though not necessarily in the way predicted by standard open system techniques. 
Thus, we apply a novel technique called the Hilbert space Average Method (HAM) and verify its results numerically 
}

\maketitle

%
%

%
%
\section{Introduction}
\label{sec:1}

Due to the linearity of the Schr\"odinger equation concepts like ergodicity or mixing are strictly speaking absent in quantum mechanics. 
Hence the tendency towards equilibrium is not easy to explain. 
However, except for some ideas \cite{Neumann1929,Landau1980} the approaches to thermalization in the quantum domain seem to be centered around the idea of a thermostat, i.e., some environmental quantum system (bath, reservoir), enforcing equilibrium upon the considered system. 
Usually it is assumed that the classical analogon of this bath contains an infinite number of decoupled degrees of freedom.

Theories addressing such scenarios are the projection operator techniques (such as Nakajima-Zwanzig or the time-convolutionless method see \cite{Breuer2002}) and the path integral technique (Feynman Vernon \cite{Weiss1999}).
The projection operator techniques are exact if all orders of the system-bath interaction strength are taken into account which is practically unfeasible. 
However, assuming weak interactions and accordingly truncating at leading order in the interaction strength, which goes by the name of ``Born approximation'' (BA), produces  an exponential relaxation behavior (cf.~\cite{Caldeira1985,Makri1999}) whenever the bath consists of an continuum of oscillators, spins, etc. 
The origin of statistical dynamics is routinely based on this scheme, if it breaks down no exponential thermalization can a priori be expected.

In contrast to the infinite baths which are extensively discussed in the mentioned literature, we will concentrate in this article on finite environments. 
The models we analyze may all be characterized by a few-level-system (S,
considered system) coupled to a many-level-system (E, environment) consisting of several relevant energy bands each featuring a number of energy eigenstates (e.g.\ see Fig.~\ref{fig:1}). 
Thus, this may be viewed as, e.g., a spin coupled to a single molecule, a one particle quantum dot, an atom or simply a single harmonic oscillator. 
Note that the spin, unlike in typical oscillator baths or the Jaynes-Cummings Model, is not supposed to be in resonance with the environments level spacing but with the energy distance between the bands. 
There are two principal differences of such a finite environment level scheme from the level scheme of, say, a standard oscillator bath: 
i) The total amount of levels within a band may be finite. 
ii) Even more important, from e.g.\ the ground state of a standard bath there are infinitely many resonant transitions to the ``one-excitation-states'' of the bath. 
But from all those, the resonant transitions lead back to only one ground state. 
Thus, the relevant bands of any infinite bath would consist of only one state in the lowest band and infinitely many states in the upper bands. 
In this paper we focus on systems featuring arbitrary numbers of states in any band. (For a treatment of finite baths under a different perspective, see \cite{Kolovsky1994,Scarani2002}).
 
It turns out that for the above mentioned class of models standard methods do not converge and thus the unjustified application of the BA produces wrong results (cf.~\cite{Breuer2006,Gemmer2006I}).
This holds true even and especially in the limit of weak coupling and arbitrarily dense environmental spectra.
Nevertheless, as the application of the Hilbert Space Average Method (HAM) predicts, a statistical relaxation behavior can be induced by finite baths.
It simply is not the behavior predicted by the BA. 
Thus, the principles of statistical mechanics in some sense apply below the infinite particle number limit and beyond the BA.
This also supports the concept of systems being driven towards equilibrium through increasing correlations with their environments \cite{Lubkin1978,Lubkin1993,Zurek1994,GemmerOtte2001,Scarani2002} rather than the idea of system and environment remaining factorizable, which is often attributed to the BA \cite{Weiss1999,Breuer2002}.

Our paper is organized as follows: 
First we introduce our class of finite environment models and the appropriate variables (\refsec{sec:2:subsec:1}, \refsec{sec:2:subsec:2}). 
Then we compute the short time dynamics of those variables (\refsec{sec:2:subsec:3}). 
Hereafter we introduce HAM and show in some detail how it can be exploited to infer the typical full time relaxation from the short time dynamics (\refsec{sec:3}). 
The theory is then verified by comparing the HAM predictions with numerically exact solutions of the time dependent Schr\"odinger equation for the respective models (\refsec{sec:4}). 
In the following Section the limits of the applicability of HAM which turn out to be the limits of the statistical relaxation itself are discussed (\refsec{sec:5}). 
Finally we conclude (\refsec{sec:6}).

%
%
\section{System and Dynamics}
\label{sec:2}

\subsection{Finite Environment Model}
\label{sec:2:subsec:1}

\begin{figure}
  \centering
  \includegraphics[width=6cm]{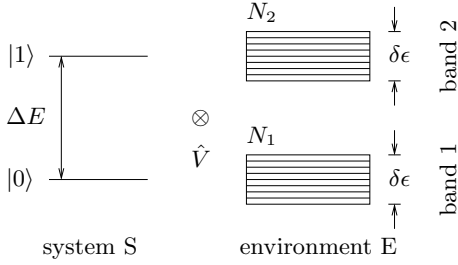}
  \caption{Two-level system coupled to a finite environment. Special case for only two bands in the environment} 
  \label{fig:1}
\end{figure}

As just mentioned we analyze a few-level-system S, with state space
${\mathcal{H}}_{\text{S}}$, coupled to a single-many-level system E
with state space ${\mathcal{H}}_{\text{E}}$ consisting of energy
bands, featuring, for simplicity, the same width and equidistant level spacing.
A simple example is depicted in Fig.~\ref{fig:1} with two bands in the environment only.
The Hilbert space of states of the composite system is given by the tensor product ${\mathcal{H}}={\mathcal{H}}_{\text{S}}\otimes{\mathcal{H}}_{\text{E}}$. 

In $\mathcal{H}_{\text{S}}$ let us introduce standard transition operators $\hat{P}_{ij}=\ketbra{i}{j}$, where $\ket{i}$, $\ket{j}$ are energy eigenstates of the considered system S.
Furthermore, we define projection operators that implement projections onto the lower, respectively upper band of the environment in $\mathcal{H}_{\text{E}}$ by
\begin{equation}
  \label{eq:1}
  \hat{\Pi}_a = \sum_{n_a} \ketbra{n_a}{n_a}\,,
\end{equation}
where $\ket{n_a}$ are energy eigenstates of E, $a$ labels the band number.
Those projectors meet the standard property 
\begin{equation}
  \label{eq:2}
  \hat{\Pi}_a\hat{\Pi}_{a'}=\delta_{aa'}\hat{\Pi}_{a'}\,.
\end{equation}
Thus, the number of eigenstates in band $a$ is given by $N_a\equiv\trtxt{\hat{\Pi}_a}$.

The complete Schr\"odinger picture Hamiltonian of the model consists of a local and an interaction part $\hat{H} = \hat{H}_{\text{loc}} + \hat{V}$, where the local part reads
\begin{equation}
  \label{eq:3}
  \hat{H}_{\text{loc}} 
  = \hat{H}_{\text{S}}\otimes\hat{1}_{\text{E}} 
   + \hat{1}_{\text{S}}\otimes\hat{H}_{\text{E}}
\end{equation}
with
\begin{align}
  \label{eq:4}
  \hat{H}_{\text{S}} 
  &= \sum_i E_i\hat{P}_{ii}\,,
  \notag\\
  \hat{H}_{\text{E}} 
  &= \sum_a \sum_{n_a=1}^{N_a} 
     \left(E_a + \frac{\delta\epsilon n_a}{N_a}\right)
     \ketbra{n_a}{n_a}\,.
\end{align}
Here we introduced the energy levels $E_i$ of S and the mean band energies $E_a$ of the environment.
Note that $\com{\hat{\Pi}_a}{\hat{H}_{\text{E}}}=0$.
For the special case depicted in Fig.~\ref{fig:1} one gets $a=1,2$ only.

\begin{figure}
  \centering
  \includegraphics[width=8cm]{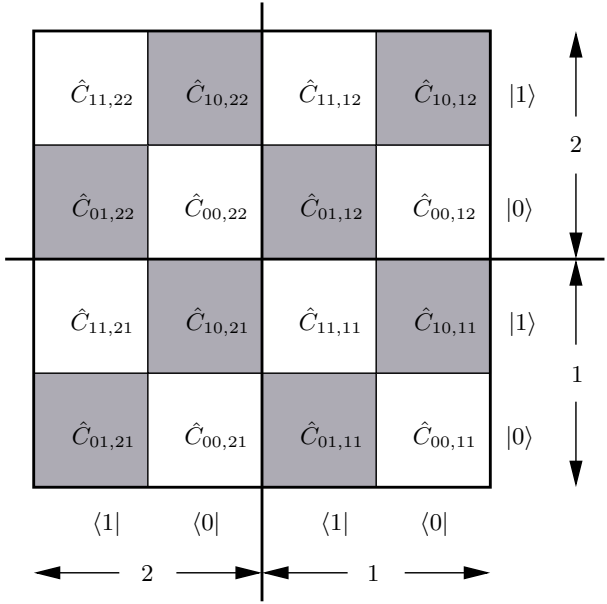}
  \caption{Scheme of the interaction matrix for the special model Fig.~\ref{fig:1}, explanation see text.}
  \label{fig:2}
\end{figure}

The interaction may, in principle, be any Hermitian matrix defined on $\mathcal{H}$. We choose to decompose it uniquely as follows
 \begin{equation}
  \label{eq:5}
  \hat{V}=\sum_{ij}\hat{P}_{ij}\otimes\hat{C}_{ij}\;,
\end{equation}
where the $\hat{C}_{ij}$ themselves may be decomposed as
\begin{equation}
  \label{eq:5a}
  \hat{C}_{ij}
  =\sum_{ab}\hat{C}_{ij,ab}
  \quad\text{with}\quad 
  \hat{C}_{ij,ab}
  =\hat{\Pi}_a \hat{C}_{ij}\hat{\Pi}_b\,. 
\end{equation}
For our special case, the interaction $\hat{V}$ and its decomposition are sketched in Fig.~\ref{fig:2}.
For later reference we define the ``coupling strengths''
$\lambda_{ij,ab}$ as
\begin{equation}
  \label{eq:6}
  \lambda_{ij,ab}^2
  := \frac{\trtxt{\hat{C}_{ij,ab}\,\hat{C}_{ji,ba}}}{N_aN_b}
\end{equation}
and due to the Hermiticity of the interaction $\lambda_{ij,ab}=\lambda_{ji,ba}$ is a real number. 
Conditions on those interaction strengths are discussed in more detail in Sec.~\ref{sec:5:subsec:1}, but in general we assume them to be weak compared to local energies in S and E, i.e., $\Delta E$ from (\ref{eq:4}).

There are two different types of (additive) contributions to $\hat{V}$: 
$\hat{C}$-terms that induce transitions inside the system S (featuring $i\neq j$) as well as terms which do not (featuring $i=j$).
Since the first type exchanges energy between system and environment it is sometimes referred to as  ``canonical coupling'' $\hat{V}_{\text{can}}$ (those terms are shaded grey in Fig.~\ref{fig:2}). 
The second type produces some entanglement thereby causing decoherence (those terms are white in Fig.~\ref{fig:2}), but does not exchange energy and therefore refers to a ``microcanonical coupling'' $\hat{V}_{\text{mic}}$ in the context of  quantum thermodynamics (cf.~\cite{Gemmer2004}).
We do not specify the interaction in more detail here. 
To keep the following theoretical considerations simple we only impose two further conditions on the interaction matrix.
First we require that the different parts of $\hat{V}$ as displayed in Fig.~\ref{fig:2} are not correlated unless they are adjoints of each other. 
Hence we get for the following traces
\begin{equation}
  \label{eq:7a}
  \trtxt{\hat{C}_{ij,ab}\,\hat{C}_{j'i',b'a'}}
  \approx N_aN_b\,\lambda_{ij,ab}^2\,
  \delta_{i,i'}\delta_{j,j'}\delta_{a,a'}\delta_{b,b'}\,.
\end{equation}
Furthermore, demand the traces over individual contributions of $\hat{V}$ to vanish, i.e.,
\begin{align}
  \label{eq:8}
\trtxt{\hat{C}_{ij,ab}}\approx 0\,.
\end{align}
Both constraints are common in the field and definitely apply for the used numerical examples.

For all numerical investigations (see \refsec{sec:4}) we are using complex Gaussian distributed random matrices with zero mean to model the interaction.
Thus, the mentioned special conditions apply: 
Only adjoint blocks are correlated and the above traces are extremely small for Gaussian random numbers with zero mean. 
This interaction type has been chosen in order to keep the model as general and free from peculiarities as possible. 
For example, in the fields of nuclear physics or quantum chaos random matrices are routinely used to model unknown interaction potentials.
We do, however, analyze the dynamics generated by one single interaction, not the average dynamics of an Gaussian ensemble of interaction matrices.

\subsection{Reduced Dynamics and Appropriate Variables}
\label{sec:2:subsec:2}

Of course we are mainly interested in the time evolution of the system S separately, i.e., we would like to find an autonomous, time-local equation for the dynamics of its reduced density matrix $\hat{\rho}$. 
However, it turns out that an autonomous description in terms of $\hat{\rho}$ is, in general, not feasible for finite environments. 
We thus aim at finding an autonomous set of equations for the dynamics of a set of variables that contain slightly more information than $\hat{\rho}$, such that from the knowledge of this set $\hat{\rho}$ may always be computed.

We simply name the set here, and explain in the following the derivations of its dynamics. 
Consider the following operators
\begin{equation} 
  \label{eq:10}
  \proj{ij}{a} = \hat{P}_{ij} \otimes \hat{\Pi}_a\,.
\end{equation}
According to (\ref{eq:2}) we find 
\begin{equation}
  \label{eq:11}
  \proj{ij}{a}\proj{i'j'}{a'}=\delta_{ji'}\delta_{aa'}\proj{ij'}{a'}\,
\end{equation}
and thus the given operators together with zero form a group which we mention here for later reference.
Throughout this paper we think of the full system as being always in a pure state $\ket{\psi}\bra{\psi}$.
Thus the expectation values of the operators (\ref{eq:10}) may be denoted as 
\begin{equation}
  \label{eq:13}
  \bra{\psi}\proj{ij}{a}\ket{\psi} \equiv P_{ij,a}\,.
\end{equation}
For the dynamics of those expectation values we are going to derive an autonomous set of equations. 
In terms of those variables the reduced density matrix elements $\rho_{ij}$ read
\begin{equation}
  \label{eq:14}
  \rho_{ij}=\sum_a P_{ij,a}\,.
\end{equation}
This may simply be computed from the definition of density matrix of S
\begin{equation}
  \label{eq:14a}
 \rho_{ij}
 =\bra{i}\hat{\rho}\ket{j}
 =\bra{i}\trtxt[\text{E}]{\ketbra{\psi}{\psi}}\ket{j}\,.
\end{equation}

\subsection{Short Time Dynamics}
\label{sec:2:subsec:3}

In order to find the full dynamics of the $P$'s defined in (\ref{eq:13}), we start of by computing their short time evolutions in this Section.
To those ends we change from the Schr\"odinger to the interaction (Dirac) picture in which the originally constant interaction $\hat{V}$ from \refsec{sec:2:subsec:1} becomes time dependent
\begin{equation}
  \label{eq:15}
  \hat{V}(t) 
  = \E^{\I\hat{H}_{\text{loc}}t/\hbar}
    \,\hat{V}\,
    \E^{-\I\hat{H}_{\text{loc}}t/\hbar}\,.
\end{equation}
As well-known in the interaction picture the time evolution may be written in terms of an propagator $\hat{D}(\tau, t)$
\begin{equation}
  \label{eq:17}
  \ket{\psi(t+\tau)}
  =\hat{D}(\tau, t)\;\ket{\psi(t)}\;,
\end{equation}
(from here $\ket{\psi(t)}$ refers to the interaction picture). 
The propagator $\hat{D}(\tau, t)$ may be explicitly noted in terms of an Dyson expansion
\begin{equation}
  \label{eq:18}
  \hat{D}(\tau,t) 
  = \hat{1}+\sum_{j=1}^{\infty}\Big(-\frac{\I}{\hbar}\Big)^j\hat{U}_j(\tau,t) 
\end{equation}
where
\begin{equation}
  \label{eq:18a}
  \hat{U}_j(\tau,t)
  =\mathcal{T}\prod_{n=1}^j\int_t^{\tau_n+t}\D\tau_n\,\hat{V}(\tau_n+t)
\end{equation}
with $\mathcal{T}$ being the standard time ordering operator.

The time evolution of the expectation values $P_{ij,a}$ according to (\ref{eq:13}) reads
\begin{equation}
  \label{eq:24}
  P_{ij,a}(t+\tau)=\bra{\psi(t+\tau)}\proj{ij}{a}\ket{\psi(t+\tau)}\,
\end{equation}
which using (\ref{eq:17}) may also be written as
\begin{equation}
  \label{eq:25}
  P_{ij,a}(t+\tau) 
  =
  \bra{\psi(t)}\proj{ij}{a}(t+\tau) \ket{\psi(t)}\,,
\end{equation}
with
\begin{equation}
  \label{eq:25a}
  \proj{ij}{a}(t+\tau) 
  =
  \hat{D}^{\dagger}(\tau,t)\,\proj{ij}{a}\,\hat{D}(\tau,t)\,.
\end{equation}
The above definition allows to write the expectation value of $P_{ij,a}$ at time $t+\tau$ (in the interaction picture) as an expectation value of some operator $\proj{ij}{a}(t+\tau)$ at time $t$. 
This particular form is well suited to asses that dynamics by HAM as will be explained in the next Section.

If $\tau$ is short and the interaction is weak the propagator may be approximated by a truncation of the Dyson series (\ref{eq:18}) to leading order which is in this case second order. 
Let the truncated propagator be denoted as $\hat{D}_2(\tau,t)$. 
This truncated propagator can typically be computed, even if complete diagonalization is far beyond reach (for a more explicit treatment of $\hat{D}_2$, cf.~App.~A). 
Thus the approximate form we are going to use in \refsec{sec:3:subsec:2} reads
\begin{equation}
  \label{eq:25b}
\proj{ij}{a}(t+\tau) 
  =
  \hat{D}_2^{\dagger}(\tau,t)\,\proj{ij}{a}\,\hat{D}_2(\tau,t)\,.
\end{equation}

%
%
\section{Dynamical Hilbert Space Average Method}
\label{sec:3}

\subsection{Definition and Calculation of the Hilbert Space Average}
\label{sec:3:subsec:1}

The Hilbert space average method (HAM) is in essence a technique to produce guesses for the values of quantities defined as functions of a wave function $\ket{\psi}$ if $\ket{\psi}$ itself is not known in full detail, only some features of it. 
In particular it produces a guess for some expectation value $\bra{\psi}\hat{S}\ket{\psi}$ [cf.~(\ref{eq:25})] if the only information about $\ket{\psi}$ is the set of expectation values $\bra{\psi}\proj{ij}{a}\ket{\psi}=P_{ij,a}$ mentioned below. 
Such a statement naturally has to be a guess since there are in general many different $\ket{\psi}$ that are in accord with the given set of $P_{ij,a}$, but produce possibly different values for $\bra{\psi}\hat{S}\ket{\psi}$. 
The question here is whether the distribution of $\bra{\psi}\hat{S}\ket{\psi}$'s produced by the respective set of $\ket{\psi}$'s is broad or whether almost all those $\ket{\psi}$'s yield $\bra{\psi}\hat{S}\ket{\psi}$'s that are approximately equal. 
It turns out that if the spectral width of $\hat{S}$ is not too large and  $\hat{S}$ is high-dimensional almost all individual $\ket{\psi}$ yield an expectation value close to the mean of the distribution of $\bra{\psi}\hat{S}\ket{\psi}$'s (see \refsec{sec:5} and \cite{Gemmer2004}). 
The occurrence of such typical values in high-dimensional systems has recently also been exploited to explain the origin of statistical behavior in \cite{Goldstein2006,Popescu2005}.

To find the above mean one has to average with respect to the $\ket{\psi}$'s. 
We call this a Hilbert space average $S$ and denote it as
\begin{equation} 
  \label{eq:26}
  S=\Haver{\bra{\psi}\hat{S}\ket{\psi}}_{\{\bra{\psi}\proj{ij}{a}\ket{\psi}=P_{ij,a}\}}
  \,.
\end{equation}
This expression stands for the average of $\bra{\psi}\hat{S}\ket{\psi}$ over all $\ket{\psi}$ that feature $\bra{\psi}\proj{ij}{a}\ket{\psi}=P_{ij,a}$ but are uniformly distributed otherwise. 
Uniformly distributed means invariant with respect to all unitary transformations $ \E^{\I\hat{G}} $ that leave the respective set of expectation values unchanged, i.e., $\bra{\psi} \E^{\I\hat{G}}\proj{ij}{a} \E^{-\I\hat{G}}  \ket{\psi}
=\bra{\psi}\proj{ij}{a}\ket{\psi}$. Thus the respective transformations may be characterized by 
\begin{equation}
  \label{eq:26a} 
  [\hat{G},\proj{ij}{a}]=0\,.
\end{equation}
Instead of computing the so defined  Hilbert space average (\ref{eq:26}) directly by integration as done, e.g.\ in \cite{Gemmer2004,Gemmer2005I} we will proceed in a slightly different way, here. 
To those ends we change from the notion of an expectation value of a state to one of a density operator
\begin{equation}
  \label{eq:27}
  S = \Haver{\bra{\psi}\hat{S}\ket{\psi}}
    = \Haver{\trtxt{\hat{S}\ketbra{\psi}{\psi}}}\,,
\end{equation}
where we skipped the constant expectation values of the Hilbert space average for the moment.
Exchanging the average and the trace, one may rewrite 
\begin{equation} 
  \label{eq:28}
  S = \trtxt{\hat{S}\Haver{\ketbra{\psi}{\psi}}}
    \equiv \trtxt{\hat{S}\hat{\alpha}}
\end{equation}
with
\begin{equation}
  \label{eq:29}
  \hat{\alpha}\equiv
  \Haver{\ketbra{\psi}{\psi}}_{\{\bra{\psi}\proj{ij}{a}\ket{\psi}=P_{ij,a}\}}
  \,.
\end{equation}
To compute $\hat{\alpha}$ we now exploit its invariance properties.
Since the set of all $\ket{\psi}$ that ``make up'' $\hat{\alpha}$ [that belong to the averaging region of (\ref{eq:29})] is characterized by being invariant under the above transformations $\E^{-\I\hat{G}}$, $\hat{\alpha}$ itself  has to be invariant under those transformations, i.e.
\begin{equation} 
  \label{eq:30}
  \E^{\I\hat{G}}\hat{\alpha}\E^{-\I\hat{G}}=\hat{\alpha} \,.
\end{equation}
This, however, can only be fulfilled if $[\hat{G},\hat{\alpha}]=0$ for all possible $\hat{G}$. 
Due to (\ref{eq:26a}) the most general form of $\hat{\alpha}$ which is consistent with the respective invariance properties is 
\begin{equation} 
  \label{eq:31}
  \hat{\alpha}=\sum_{ija} p_{ij,a} \proj{ij}{a}\,,
\end{equation}
where the coefficients $p_{ij,a}$ are still to be determined.
In principle the above sum could contain addends of higher oder, i.e., products of the $\hat{P}$-operators, but according to the properties of the projection and transition operators [especially (\ref{eq:11})], those products reduce to a single $\hat{P}$-op\-er\-a\-tor or zero (in other words, the $\proj{ij}{a}$ form a group), hence (\ref{eq:31}) is indeed the most general form.  

How are the coefficients $p_{ij,a}$ to be determined? 
From the definition of $\hat{\alpha}$ in (\ref{eq:29}) it follows
\begin{equation} 
  \label{eq:31a}
  \trtxt{\hat{\alpha}\proj{i'j'}{a'}} = P_{i'j',a'}\,.
\end{equation}
By inserting (\ref{eq:31}) into (\ref{eq:31a}) and exploiting (\ref{eq:11}) the coefficients are straightforward found to be
\begin{equation}
  \label{eq:33}
  p_{ij,a} = \frac{P_{ji,a}}{N_a}\,.
\end{equation}
Thus, we finally get for the Hilbert space average (\ref{eq:28})
\begin{equation}
  \label{eq:34}
  S = \trtxt{\hat{S}\hat{\alpha}}
    = \sum_{ija} \frac{P_{ji,a}}{N_a}\trtxt{\hat{S}\proj{ij}{a}}\,.
\end{equation}

\subsection{Iterative Guessing}
\label{sec:3:subsec:2}

\begin{figure}
  \centering
  \includegraphics[width=6.5cm]{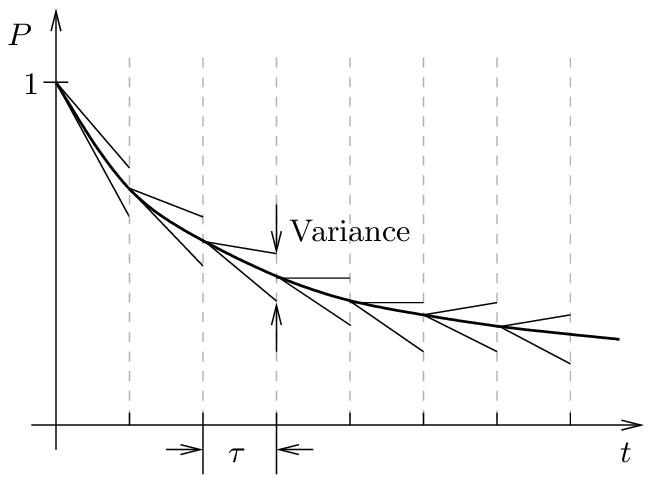}
  \caption{Repeated guessing scheme. To each iteration time step
    $\tau$ corresponds an increasing uncertainty (variance) of the guess.}
  \label{fig:3}
\end{figure}

To find the (reduced) autonomous dynamics for the $P_{ji,a}$ from HAM we employ the following scheme: 
Based on HAM we compute a guess for the most likely value of the set $P_{ji,a}$ at time $(t+\tau)$ [i.e.,$P_{ji,a}(t+\tau)$] assuming that we knew the values for the $P_{ji,a}$ at time $t$ [i.e.,$P_{ji,a}(t)$]. 
Once such a map $P_{ji,a}(t) \rightarrow P_{ji,a}(t+\tau)$ is established it can of course be iterated to produce the full time dynamics. 
This of course implies repeated guessing, since in each iteration step the guess from the step before has to be taken for granted. 
However, if each single guess is sufficiently reliable, i.e., the spectrum of possible outcomes is rather sharply concentrated around the most frequent one (which one guesses), even repeated guessing may yield a good ``total'' guess for the full time evolution. 
The scheme is schematically sketched in \reffig{fig:3}.
Some information about the reliability of HAM guesses has already been given in \refsec{sec:3:subsec:1}, the applicability of the whole scheme will be analyzed more thoroughly in \refsec{sec:5}. 

To implement the above scheme we consider the equation one gets from inserting $\proj{ij}{a}(t+\tau)$ for $\hat{S}$ in (\ref{eq:34})
\begin{align} 
 \label{eq:35}
&\Haver{\bra{\psi}\proj{ij}{a}(t+\tau)\ket{\psi}}_{\{\bra{\psi}\proj{i'j'}{a'}\ket{\psi}=P_{i'j',a'}(t)\}} \notag\\
& =\sum_{i'j'a'} \frac{P_{j'i',a'}(t)}{N_{a'}}
  \trtxt{\hat{D}_2^{\dagger}(\tau,t)\proj{ij}{a}\hat{D}_2(\tau,t)\proj{i'j'}{a'}}\,.
\end{align}
This is the Hilbert space average (HA) over all possible $P_{ji,a}(t+\tau)$ under the condition that one had at time $t$ the set $P_{ji,a}(t)$. 
Thus the (iterative) guess now simply consists of replacing the HA by the actual value, i.e., 
\begin{equation}
  \label{eq:36}
  P_{ij,a}(t+\tau)
  \approx
  \sum_{i'j'a'} \frac{P_{j'i',a'}(t)}{N_{a'}}
  \trtxt{\hat{D}_2^{\dagger}\proj{ij}{a}\hat{D}_2\proj{i'j'}{a'}}\,.
\end{equation}
The evaluation of the right hand side requires some rather lengthy calculations, but can be done without further assumptions or approximations. 
The interested reader may find the details in App.~B. 
Here we simply give the results and proceed.

For $P$'s featuring $i=j$ one finds
\begin{align}
  \label{eq:58}
  &P_{ii,a}(t + \tau)-P_{ii,a}(t)\notag\\
  &\quad=\sum_{mb}2\mbox{Re}f_{im,ab}(\tau)
         \Big(\frac{P_{mm,b}(t)}{N_b}- \frac{P_{ii,a}(t)}{N_a}\Big)
\end{align}
and for the $P$'s with $i\neq j$
\begin{align} 
  \label{eq:59}
  &P_{ij,a}(t+\tau)-P_{ij,a}(t)
  \notag\\
  &\quad=-\frac{1}{2}\frac{P_{ij,a}(t)}{N_a}
         \sum_{mb} \big(f_{jm,ab}(\tau)+f^*_{im,ab}(\tau)\big)\,,
\end{align}
where the $f(\tau)$'s are defined as
\begin{align}
  \label{eq:39}
  f_{ij,ab}(\tau)
  &:= \int_0^{\tau}\D\tau'\int_0^{\tau'}\D\tau''  g_{ij,ab}(\tau'')\,\\
  \label{eq:7}
  g_{ij,ab}(\tau'') 
  &:=\frac{1}{\hbar^2}\,
    \trtxt[\text{E}]{\hat{C}_{ab,ij}(\tau'')\,\hat{C}_{ba,ji}}
     \,.
\end{align}
Note that (\ref{eq:58}) and (\ref{eq:59}) now are autonomous (closed) in terms of the respective $P$'s and there is no more explicit dependence on the absolute time $t$. 
It turns out (see below) that the  $f(\tau)$'s are approximately linear in $\tau$. 
Hence for the squared absolute values of the $P$'s with $i \neq j$ (which we will be primarily analyzing rather than the $P$'s themselves) one finds to linear order in $\tau$
\begin{align} 
  \label{eq:59d}
  &|P_{ij,a}(t+\tau)|^2-|P_{ij,a}(t)|^2
  \notag\\
  &=
  -\frac{|P_{ij,a}(t)|^2}{N_a}
  \sum_{mb} \big(\mbox{Re}f_{jm,ab}(\tau)+\mbox{Re}f_{im,ab}(\tau)\big)\,.
\end{align}

\subsection{Correlation Functions and Transition Rates}
\label{sec:3:subsec:3}

In order to interpret (\ref{eq:58}) and (\ref{eq:59d}) appropriately, we  need some information about the correlation functions $f(\tau)$. 
Apparently those $f(\tau)$'s are essentially integrals over the same environmental temporal correlation functions $g(\tau'')$ that appear in the memory kernels of standard projection operator techniques. 
(Only here they explicitly correspond to transitions between different energy subspaces of the environment.) 
Thus we analyze the $g(\tau'')$'s from (\ref{eq:39}) more thoroughly. 
Their real parts (which eventually essentially matter) read
\begin{align} 
  \label{eq:59a}
  \text{Re}\,g_{ij,ab}(\tau'')
  =& \frac{1}{\hbar^2}\sum_{n_a,n_b}
     |\bra{n_a}\hat{C}_{ij}\ket{n_b}|^2 \times
  \\ 
   & \cos\left(|E_i-E_j+E(n_a)-E(n_b)|\frac{\tau''}{\hbar}\right)\,.\notag
\end{align}
Thus they simply consist of a  sum of weighted cosine functions with different frequencies. 
The set of those weights essentially gives the Fourier transform of the corresponding correlation function. 
First of all, only if the transition within the system ($j \rightarrow i$) is in resonance with the energy gap between the bands $a,b$, $g(\tau)$ will contain any small frequency contributions at all. 
Hence, only in this case temporal integrations, i.e., the corresponding $f$'s will be nonzero. 
In the resonant case the frequency spectrum will  stretch from zero to a frequency on the order of $\delta \epsilon/\hbar$, at least if the interaction gives rise to the corresponding transitions of the environment. 
Thus $g(\tau)$ will decay on a timescale on the order of $\tau_c$ with  
\begin{equation}
  \label{eq:59b}
  \tau_c \approx  \frac{\hbar}{\delta \epsilon}\,.
\end{equation}
For $\tau > \tau_c$, $g(\tau)$ will be essentially zero. 
This means that $f(\tau)$ which is a twofold temporal integration of $g(\tau)$ will grow linear in time, i.e., $f(\tau)=\gamma\tau$ after $\tau \approx\tau_c$.
The factor $\gamma$ is given by the area under the curve  $g(\tau)$ up to approximately $\tau_c$. 
If $\tau_c$ was infinite $\gamma$ would only be determined by the weight of the zero-frequency terms of $g(\tau)$. 
Since $\tau_c$ is finite, $\gamma$ is related to the ``peak density'' of $g$ within frequency range from zero to $\Delta \omega$ with $\Delta \omega_c \ll   1/\tau_c \approx \delta \epsilon/\hbar$. 
Which means $\gamma$ is eventually  given by the sum of all weights that correspond to frequencies from zero to $\Delta \omega$ divided by $\Delta \omega$ and multiplied by $\pi$. 
Since in our model the $|\bra{n_a}\hat{C}_{ij}\ket{n_b}|$ are Gaussian distributed random numbers we eventually find for the  $f(\tau)$'s 
\begin{equation}
  \label{eq:59c}
  \text{Re}\,f_{ij,ab}(\tau)
  \approx \frac{\pi\lambda_{ij,ab}^2N_aN_b}{\hbar\delta\epsilon}\;\tau\,.
\end{equation}
This result can apparently be connected to the transition rate as obtained from Fermi's Golden Rule. 
Let $\gamma_{ij,ab}$ be the Golden Rule transition rate for a transition of full system characterized by $j \rightarrow i$ and $b \rightarrow a$. 
Then the connection reads
\begin{equation} 
  \label{eq:60}
  2\,\text{Re}\,f_{im,ab}(\tau)\approx \gamma_{im,ab}\, N_b\;\tau\,.
\end{equation}
Since the Golden Rule transition rates depend on the state densities around the final states, respective ``forward'' and ``backward'' rates are, for equal bandwidths, connected as 
\begin{equation}
  \label{eq:61}
  \gamma_{im,ab}=\frac{N_a}{N_b}\,\gamma_{mi,ba}\;.
\end{equation}

\subsection{Reduced Equations of Motion}
\label{sec:3:subsec:4}

Inserting  (\ref{eq:60}) and (\ref{eq:61}) into (\ref{eq:58}) and (\ref{eq:59d}) allows for a computation of the full dynamics of the $P_{ii,a}$ and the $|P_{ij,a}|^2$ through iteration. 
The iteration has to proceed in time-steps that are longer than $\tau_c$, but shorter than  $\tau_d$. 
The latter will be explained in \refsec{sec:5:subsec:2}. 
Assuming that $\tau_c$ is short compared to the timescale of the relaxation dynamics it may be written as 
\begin{align} 
  \label{eq:62}
  &\dod{}{t}P_{ii,a}(t)
  = \sum_{mb}
     \big[\gamma_{im,ab}P_{mm,b}(t)-\gamma_{mi,ba}P_{ii,a}(t)\big]\,, 
  \\
  \label{eq:63}
  &\dod{}{t}|P_{ij,a}(t)|^2
  = - |P_{ij,a}(t)|^2
  \sum_{mb}(\gamma_{mi,ba}+\gamma_{mj,ba})\,.
\end{align}
(This form is in accord with recent results from novel projection
operator techniques \cite{Breuer2006})
We now analyze this set of equations in a little more detail. 
The $P_{ii,a}$ may be interpreted as the probability to find the joint system in state $i$ for S and in band $a$ with respect to the environment. 
Equation (\ref{eq:62}) obviously has the form of a master equation, i.e, the overall probability is conserved and there is a stable fixpoint which sets the equilibrium values for the $P_{ii,a}$. 
According to (\ref{eq:63}) the $P_{ij,a}$ will all decay to zero. 
Taking (\ref{eq:14}) into account this implies that $\hat{\rho}$ will reach an equilibrium state which is diagonal in the basis of the energy eigenstates of S. 
As already mentioned below (\ref{eq:59a}) transitions occur only between resonant states, i.e., the $\gamma_{im,ab}$ are zero unless $E_m+E_b\approx E_i+E_a$ where $E_a,E_b$ are the corresponding mean band energies. 
Thus, if we define the approximate full energy of some state $E$
\begin{equation}
  \label{eq:63a}
  E \equiv E_m+E_b\,,\quad E \equiv E_i+E_a\,,
\end{equation}
we may label full system states by $i,E(m,E)$ rather than $i,a(m,b)$ and nonzero transition rates by $im,E$ rather than $im,ab$, i.e, $P_{ii,a}\rightarrow P_i^E$, $\gamma_{im,ab} \rightarrow\gamma_{im}^E$.
With this index transformation and exploiting (\ref{eq:61}) we may rewrite (\ref{eq:62}) as 
\begin{equation}
  \label{eq:63b}
  \dod{}{t}P_i^E(t)
  = \sum_{m}
    \big[\gamma_{im}^E P_{m}^E(t)-\frac{N(E-E_m)}{N(E-E_i)}
         \gamma_{im}^E P_{i}^E(t)\big]\,, 
\end{equation}
where $N(E-E_m)$ is the dimension of the environmental band with energy $E_b=E-E_m$. 
This form reflects the fact that the dynamics of the occupation probabilities with different overall energies are decoupled. 
We, furthermore, find from  (\ref{eq:63b}) for the equilibrium values $P_i^E(t \rightarrow \infty)\propto N(E-E_i)$. 
Thus, the equilibrium state is in accord with the a priori postulate in that sense that the probability to find the full system in some subspace is proportional to the dimension of this subspace. 
However, it is in general impossible to transform  (\ref{eq:63b}) in a closed set of equations for the occupation probabilities $\rho_{ii}=\sum_E P_i^E$ of S alone. 
This may only be done if either only one energy subspace $E$ is occupied at all, or if the transition rates $\gamma_{im}^E$ are independent of $E$ and the number of states of the environmental bands $N_a$ scales as $N_a\propto\exp(\beta E_a)$. 
Then (\ref{eq:63b}) may be summed over $E$ yielding
\begin{equation}
  \label{eq:63c}
  \dod{}{t}\rho_{ii}(t) 
  = \sum_{m}
    \big[\gamma_{im}\rho_{mm}(t)-\E^{\beta(E_i-E_m)}
         \gamma_{im}\rho_{ii}(t)\big]\, 
\end{equation}
which is the usual closed form for the dynamics of the $\rho_{ii}$ with the standard canonical equilibrium state $\rho_{ii}(t \rightarrow\infty)\propto \exp(-\beta E_i)$. 
Thus it is essentially the exponentially growing density of states of typical infinite environments that allows for a closed dynamical description of the considered system S alone and produces the standard Gibbsian equilibrium state.%

\subsection{Thermalization and Decoherence}
\label{sec:3:subsec:5}

\begin{figure}
  \centering
  \includegraphics[width=6.5cm]{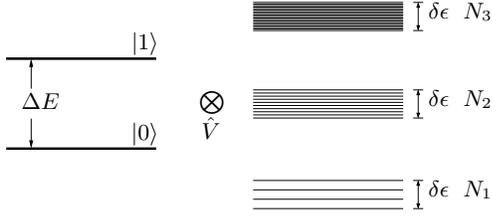}
  \caption{Three band model for the investigation of the relation
    between decoherence and thermalization}
  \label{fig:4}
\end{figure}

\begin{figure}
  \centering
  \includegraphics[width=6.5cm]{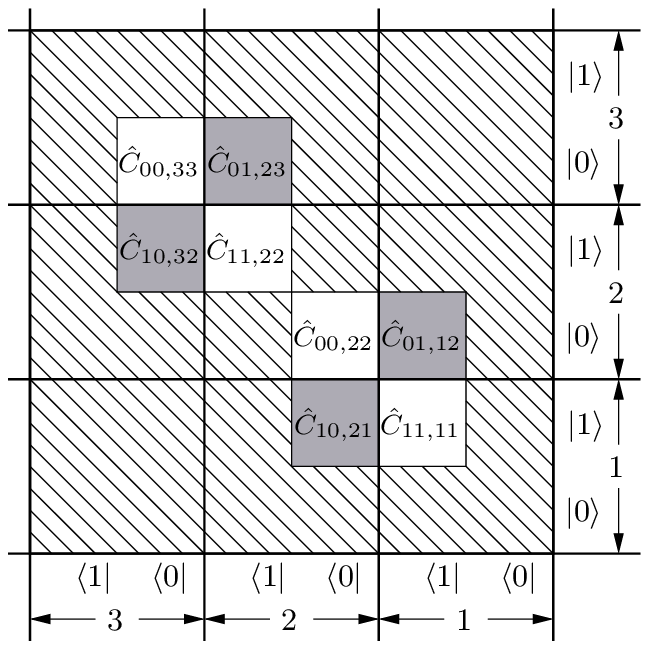}
  \caption{Interaction matrix for the model \reffig{fig:4}, canonical
    blocks (gray), microcanonical blocks (white), in case of weak
    coupling irrelevant blocks (hatched).}
  \label{fig:5}
\end{figure}

In order to investigate the relation between the decay of diagonal and
off-diagonal elements of the reduced density operator of S, we
concretely analyze as an example a slightly modified model featuring
the above mentioned structure (exponential state density, equal rates) yielding autonomous dynamics for
$\hat{\rho}$. The model is depicted in \reffig{fig:4}. 
For simplicity we consider only three environmental bands with the same density of states, i.e.,  the exponential prefactor from (\ref{eq:63c}) vanishes, $\beta =0$. 
(This eventually implies infinite temperature.) 
As mentioned the rates that control the dynamics of the diagonal elements (canonical dynamics, thermalization) have to be equal, thus, we choose $\lambda_{01,12}=\lambda_{01,23}=\lambda_{\text{can}}$. 
The rates that control the dynamics of the off-diagonal elements (microcanonical dynamics, decoherence) also have to be equal among themselves, but may differ from the ``canonical rates''.
Thus, we choose $\lambda_{00,33}=\lambda_{00,22}=\lambda_{11,22}=\lambda_{11,11}=\lambda_{\text{mic}}$.
Since all other parts of the interaction would not fulfill the resonance condition anyway, we set them to zero (cf.~Fig.~\ref{fig:5}).

Plugging those model parameters into (\ref{eq:63c}) yields
\begin{align}
  \label{eq:568}
  \dod{\rho_{00}}{t}
  &= \frac{2\pi\lambda_{\text{can}}^2 N}{\hbar\delta\epsilon}
     \big(\rho_{11} - \rho_{00}\big)\,,\\
  \label{eq:196}
  \dod{\rho_{11}}{t}
  &= \frac{2\pi\lambda_{\text{can}}^2 N}{\hbar\delta\epsilon}
     \big(\rho_{00} - \rho_{11}\big)\,.
\end{align}
Defining the thermalization time as
\begin{equation}
  \label{eq:348}
  T_{\text{th}}=\frac{\hbar\delta\epsilon}{4\pi\lambda_{\text{can}}^2 N}\,,
\end{equation}
the solution of the above set of differential equations is just an exponential decay according to $\E^{-t/T_{\text{th}}}$.

Apparently the $P_{ij,a}$ do not ``mix'' with respect to different $a$ [cf.~(\ref{eq:63})]. 
Thus, if initially the environment only occupies, e.g., band 2, for the full dynamics the off-diagonal element of $\hat{\rho}$ will be simply given by $\rho_{ij}=P_{ij,2}$ [cf.~(\ref{eq:14})]. 
In this case we find from (\ref{eq:63})
\begin{equation}
  \label{eq:350}
  \dod{|\rho_{10}|}{t}
  = - \frac{2\pi N(\lambda_{\text{can}}^2+\lambda_{\text{mic}}^2)}
           {\hbar\delta\epsilon} |\rho_{10}|\,.
\end{equation}
Using the definition $\xi=\lambda_{\text{mic}}/\lambda_{\text{can}}$, we find for the decoherence time, i.e., the time-scale on which $|\rho_{10}|$ decays
\begin{align}
  \label{eq:351}
  T_{\text{dec}}&
  = \frac{\hbar\delta\epsilon}
         {2\pi N(\lambda_{\text{can}}^2+\lambda_{\text{mic}}^2)}
  = \frac{\hbar\delta\epsilon}
         {2\pi N \lambda_{\text{can}}^2(1+\xi^2)}
  \notag\\
  &= \frac{2T_{\text{th}}}{1+\xi^2}\,.
\end{align}
For the absence of microcanonical coupling terms ($\lambda_{\text{mic}}=0\rightarrow \xi=0$) we get $2 T_{\text{th}} =T_{\text{dec}}$ which is a standard result in the context of atomic decay, quantum optics, etc. 
Nevertheless, for increasing $\xi$ decoherence may become arbitrarily faster than thermalization which is a central feature of models that are supposed to describe the motion of particles subject to heat baths, like, e.g., the Caldeira Legget model. 
Thus our model exhibits a continuous transition between those archetypes of behavior.

%
%
\section{Application}
\label{sec:4}

\subsection{Relaxation Dynamics in Model Systems}
\label{sec:4:subsec:1}

In this Section concrete models are introduced and the corresponding time dependent Schr\"odinger equations are solved. 
Then the results are compared to predictions from HAM and standard open system methods.
Our first model is of the type depicted in \reffig{fig:1}. 
The two level system features a splitting of $\Delta E=25u$.
Here and in the following we use an arbitrary energy unit $u$.
The environment consist of two bands of width $\delta \epsilon=0.5u$ with the same amount of levels $N=N_1=N_2=500$ in each one and separated also by $\Delta E=25u$. 
As already mentioned we use complex Gaussian random matrices to model the coupling, thus, satisfying the criteria \refsec{sec:2:subsec:1}.
First, we choose only a canonical interaction due to the coupling strength $\lambda_{\text{can}}=5 \cdot 10^{-4}u$ ($\lambda_{\text{mic}}=0$).  

\begin{figure}
  \centering
  \includegraphics[width=6.5cm]{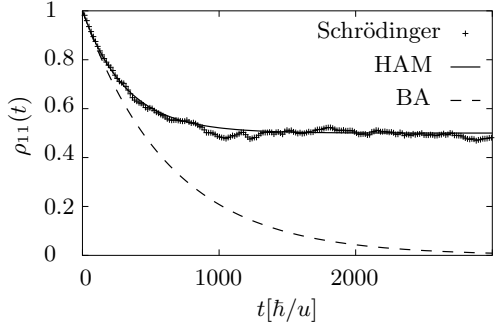}
  \caption{Evolution of the excitation probability of a product initial state. Dashed line refers to the standard master equation in born approximation.}
  \label{fig:6}
\end{figure}

\begin{figure}
  \centering
  \includegraphics[width=6.5cm]{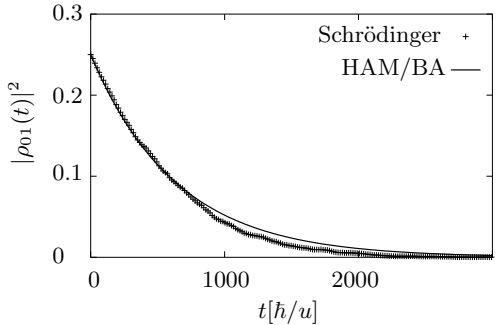}
  \caption{Off-diagonal element for a correlated initial state.}
  \label{fig:7}
\end{figure}

At first we analyze the decay behavior of two different pure product initial states. 
The environmental part of both initial states is a pure state that only occupies the lower band, but is apart from that chosen at random.
Irrespective of its pureness only with respect to occupation numbers, E's initial state can be considered an approximation to a Gibbs state with $\delta \epsilon\ll kT_{\text{E}} \ll \Delta E$ and the temperature of the environment $T_{\text{E}}$ (in the example at hand, e.g., $ kT_{\text{E}} \approx 5u$).
For small $\delta\epsilon$ the temperature may be arbitrarily small. 
Initially, the system S is firstly chosen to be completely in its excited state and, secondly, in a 50:50 superposition of ground and excited state.
The probability [density matrix element $\rho_{11}(t)$] to find the system excited as produced by the first initial state is shown in Fig.~\ref{fig:6}.
Since the first initial state does not contain any off-diagonal elements, we find $|\rho_{01}|^2\approx 0$ for all times.
This is different for the second initial state investigated in Fig.~\ref{fig:7}, it starts with $|\rho_{01}|^2=0.25$ and is thus well suited to study the decay of the coherence.
(The diagonal elements of the second state are already at their equilibrium value $\rho_{11}(0)=0.5$ in the beginning and exhibits no further change.)

By numerically solving the time-dependent Schr\"o\-din\-ger equation for the full model's pure state we find for the reduced state of the system, an  exponential decay, up to some fluctuations as depicted in \reffig{fig:6} and \reffig{fig:7}.
(For the baths initial state being a real mixed Gibbs state one can even expect fluctuations to be smaller, since fluctuations corresponding to various pure addends of the Gibbs state will partially cancel each other.) 
The solid lines are the HAM results as computed from (\ref{eq:62}) and (\ref{eq:63}). 
Obviously, they are in accord with the exact result.

The full model is Markovian in the sense that bath correlations decay much faster than the system relaxes, concretely bath correlations decay on a time scale of $\tau_c \approx \hbar /\delta \epsilon =2$ (all times given in units of $\hbar/u$), whereas the system relaxes on a timescale $T_1 \approx 640$ (cf.\ Fig. \ref{fig:6}). 
Nevertheless, S's excitation probability deviates significantly from what the standard methods (BA) predicts (cf.\ Fig.~\ref{fig:6}): 
The beginning is described correctly, but rather than ending up at temperature $T=T_\text{E}$ as the BA  predicts for thermal environment states \cite{Breuer2002}, S ends up at temperature $T=\infty$, i.e., equal occupation probabilities for both levels.
The equilibrium value of S's excitation probability is given by $\rho_{11}(\infty)=N_1/(N_1+N_2)$.
Thus, only if $N_2 \gg N_1$ (infinite bath) the BA produces correct results.
Note, however, that it is not the finite density of states that causes the break down of the BA, since the BA produces wrong results even for $N_1$, $N_2 \rightarrow \infty$ as long as the above condition is not met.
  
Furthermore, a condition often attributed to the BA, namely that S and E remain unentangled, is not fulfilled: 
When S has reached equilibrium the full system is in a superposition of $|$S in the excited state $\otimes$ E in the lower band$\rangle$ and  $|$S in the ground state $\otimes$ E in the upper band$\rangle$. 
This is a maximum entangled state with two orthogonal addends, one of which features a bath population corresponding to $T_{\text{E}}\approx 0$, the other a bath population inversion, i.e., even a negative bath temperature. 
These findings contradict the concept of factorizability, nevertheless, HAM predicts the dynamics correctly. 
This is in accord with a result from \cite{Gemmer2005I,Gemmer2006I,Breuer2006} claiming that an evolution towards local equilibrium is always accompanied by an increase of system-bath correlations. 
However, the off-diagonal element evolution coincides with the behavior predicted by the BA. 
Thus, in spite of the systems finiteness and the reversibility of the underlying Schr\"odinger equation S evolves towards maximum local von Neumann entropy (see Fig.~\ref{fig:6} and Fig.~\ref{fig:7}) which supports the concepts of \cite{Lubkin1993}.

\begin{figure}
  \centering
  \includegraphics[width=6.5cm]{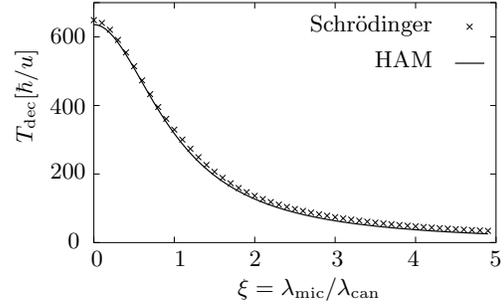}
  \caption{Dependence of the decoherence time on the microcanonical interaction strength. HAM theory according to (\ref{eq:351}).}
  \label{fig:8}
\end{figure}

To show that it is indeed possible to get different time scales for
the decay of diagonal elements of the density matrix (thermalization)
and the decay of off-diagonal elements (decoherence) according to pure
Schr\"odingerian dynamics we consider the concrete model as addressed
in \refsec{sec:3:subsec:5} with parameters $N=500$, $\delta \epsilon=0.5u$, $\Delta E=25u$ and $\lambda_{\text{can}}=5 \cdot 10^{-4}u$.
However, we choose the microcanonical interaction strength $\lambda_{\text{mic}}$, in units of the canonical one between $\xi=0$ and $\xi=5$.
As an initial state we prepared a 90:10 superposition of ground and excited state in the system, environment somewhere in the middle band.
This refers to a finite off-diagonal element in the beginning.
We have computed the Schr\"odinger dynamics of both diagonal and off-diagonal elements of the two level system.
By fitting an exponential to the off-diagonal element we get the decoherence time $T_{\text{dec}}$ in dependence of the microcanonical coupling strength.
In \reffig{fig:8} we show this numerical decoherence time $T_{\text{dec}}$ in comparison with the theoretical prediction of the HAM theory, thus (\ref{eq:351}).
As can be seen, the numerical result is in very good accordance with our theory.

\subsection{Accuracy of HAM}
\label{sec:4:subsec:2}

\begin{figure}
  \centering
  \includegraphics[width=6.5cm]{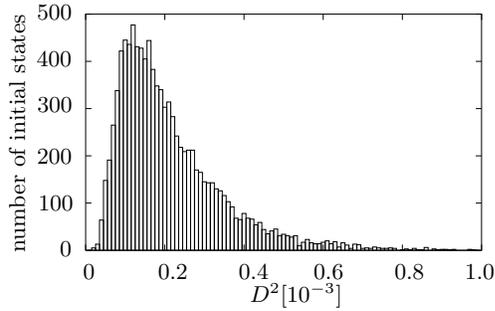}
\caption{Deviation of the exact evolution of the spins excitation probability from the HAM prediction for a set of entangled initial states.}
  \label{fig:9}
\end{figure}

Since HAM is just a ``best guess theory'' the exact evolution follows its predictions with different accuracies for different initial states, even if all conditions on the model are fulfilled.
To analyze this for, say $\rho_{11}(t)$, we introduce $D^2$, being the time-averaged quadratic deviation of HAM from the exact (Schr\"odinger) result 
\begin{equation}
  \label{eq:85}  
  D^2=\frac{1}{\nu T_1}\int_0^{\nu T_1}\D t\,
  \Big(\rho_{11}^{\text{HAM}}(t)-\rho_{11}^{\text{exact}}(t)\Big)^2\;.
\end{equation}
Thus, $D$ is a measure of the deviations from a predicted behavior.
The results of the investigation for our model (Fig.~\ref{fig:1}) are condensed in the histogram (Fig.~\ref{fig:9}, $\nu=3$, $N=500$).
The set of respective initial states is characterized by a probability of $3/4$ for $|$S in its excited state $\otimes$ E in its lower band$\rangle$ and $1/4$ for $|$S in its ground state $\otimes$ E in its upper band$\rangle$. 
Within these restrictions the initial states are uniformly distributed in the corresponding Hilbert subspace.
Since all of them are correlated the application of a product projection operator technique would practically be unfeasible. 
However, as Fig.~\ref{fig:9} shows, the vast majority of them follows the HAM prediction quite closely, although there is a  typical fluctuation of $D=\sqrt{2}\cdot 10^{-2}$ which is small compared to the features of the predicted behavior (which are on the order of one), due to the finite size of the environment (cf.\ also fluctuations in Fig.~\ref{fig:6}).

\begin{figure}
  \centering
  \includegraphics[width=6.5cm]{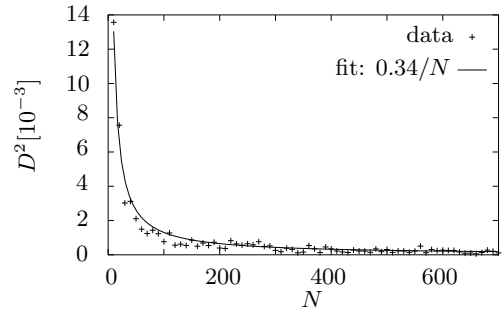}
\caption{Deviation of the exact evolution of the spins excitation probability from the HAM prediction for increasing number $N$ of states in the environment.}
  \label{fig:10}
\end{figure}

In Fig.~\ref{fig:10} the dependence of $D^2$ on the number of states of E is displayed for $N=10,\dots,800$ (one evolution for each size of the environment).
At $N=500$ like used in the above accuracy investigation we find the same typical fluctuation, whereas for smaller environments the typical deviation is much bigger.
We find that the squared deviation scales as $1/N$ with the size of the environment, thus, making HAM a reasonably reliable guess for many-state environments.

%
%
\section{Limits for the Applicability}
\label{sec:5}

The dynamical considerations of \refsec{sec:3} are only guesses, but as guesses they are valid for any initial state regardless of whether it is pure, correlated, entangled, etc. 
Thus, in contrast to the standard Nakajima-Zwanzig and TCL methods HAM allows for a direct prediction of the typical behavior of the system. 
Nevertheless, for deriving the above HAM rate equations we have claimed (and already discussed) that there is a reasonably well defined correlation time $\tau_c$ (cf.\ \refsec{sec:3:subsec:2}) at all. 
Additionally, we used two further approximations:
The truncation of the Dyson series in second order [see (\ref{eq:25b})] and the replacement of the actual value of an expectation value by the average in the respective Hilbert space compartment [see (\ref{eq:35})]. 
In the following we will investigate the validity of these approximations in more detail.

\subsection{Truncation of the Dyson Series}
\label{sec:5:subsec:1}

In (\ref{eq:25b}) we truncated the Dyson series arguing that for short times $\tau$ and small interaction strength this can be a reasonable approximation. We require, however, $\tau>\tau_c$. 
Thus, for given interaction strength, the time for which the truncation should hold, $\tau_d$  should exceed the correlation time, i.e., $\tau_d>\tau_c$. 
How can $\tau_d$ be at least approximately determined?

Consider the deviation $\ket{\delta \psi (t,\tau)}$ of a state at time $t+\tau$ from the state at time $t$, i.e., $\ket{\delta \psi (t,\tau)}:=\ket{\psi (t+\tau)}-\ket{\psi (t)}$ and let the norm of this deviation be denoted as $\Delta (t,\tau)=\sprod{\delta \psi(t,\tau)}{\delta \psi (t,\tau)}$. 
If we now evaluate  $\Delta(t,\tau)$ by means of a truncated Dyson series and find it small compared to one it is consistent to assume that higher orders are negligible for the description of $\ket{\psi (t+\tau)}$. 
If we, in contrary, find it to be large compared to one, the truncation is definitely not justified. 
Thus, we implicitly define $\tau_d$ roughly as $\Delta(t,\tau_d) \approx 1$.

Truncating the Dyson series to leading order yields (cf.\ App.~A)
\begin{equation}
  \label{eq:74}
  \Delta (t,\tau)= \bra{\psi(t)}\hat{U}_1^2 (t,\tau)\ket{\psi(t)} \,. 
\end{equation}
Since we in general do not know $\ket{\psi(t)}$ in detail, but only the $P's$ we replace, following again the argument in \refsec{sec:3:subsec:1}, the actual value of $\Delta (t,\tau)$ by its Hilbert space average $\Haver{\Delta(t,\tau)}_{\{\bra{\psi}\proj{ij}{a}\ket{\psi}=P_{ij,a}(t)\}}$, thus, obtaining
\begin{equation}
  \label{eq:75}
  \Delta (t,\tau)\approx\sum_{ija}\frac{P_{ij,a}}{N_a}\,
  \trtxt{\proj{ij}{a}(t)\,\hat{U}_1^2 (t,\tau)}\,.
\end{equation}
Exploiting (\ref{eq:55}) we find
\begin{equation}
  \label{eq:76}
  \Delta (t,\tau)=\sum_{imab}\frac{P_{ii,a}}{N_a}\,2\text{Re}\,f_{im,ab}
\end{equation}
which, taking $(\ref{eq:60})$ and  $(\ref{eq:61})$ into account and for times $\tau>\tau_c$ eventually yields
\begin{equation}
  \label{eq:77}
  \Delta (t,\tau)
  =\sum_{imab}P_{ii,a}\gamma_{mi,ba}\tau 
  =\sum_E\sum_{mi}P_i^E(t)\gamma_{mi}^E\tau\,,
\end{equation}
where the second form refers to the notation introduced in and below (\ref{eq:63a}). 
Thus, $\Delta (t,\tau)$ grows linear in $\tau$. 
Since all the probabilities $P_i^E(t)$ sum up to one at all times the growth is essentially determined by the rates $\gamma_{mi}^E$. 
Since already the sum of the $P_i^E(t)$ over $i$ and some fixed overall-energy is a constant of motion, rates belonging to energy subspaces $E$ which are not occupied in the beginning will never influence $\Delta(t,\tau)$.
Hence one should consider $\tau_d^E$, the time for which the truncation of the Dyson series holds within the invariant energy subspace $E$. From (\ref{eq:77}) we find as an rough estimate for $\tau_d^E$ 
\begin{equation}
  \label{eq:78}
  \tau_d^E\approx N_{\text{S}}\left(\sum_{mi}\gamma_{mi}^E\right)^{-1}\,,
\end{equation}
where $N_{\text{S}}$ is the number of eigenstates of S. 
Comparing this to (\ref{eq:62}) and (\ref{eq:63}) it becomes obvious that this is also roughly the time-scale for the relaxation dynamics of the $P$'s. 
This implies that the claim $\tau_d^E>\tau_c$, which guarantees the applicability of the truncation of the Dyson series, is equivalent to claiming that the typical relaxation time of the $P$'s should be long compared to the typical correlation time $\tau_c$. 
The latter has already been claimed before (\ref{eq:6}) in order to transform the iteration scheme into a differential equation. This condition can easily be controlled by changing the overall interaction strength $\lambda$. We find that for values of $\lambda$ that violate the above condition the agreement between the numerical solution and the HAM prediction vanishes.

\subsection{Hilbert Space Variance}
\label{sec:5:subsec:2}

Here we quite briefly consider the assumption that gave raise to the replacement of actual expectation values by their Hilbert space averages in \refsec{sec:3:subsec:1}. 
As already mentioned, such a replacement can only yield a reasonable result if the largest part of the possible expectation values is indeed close to the corresponding Hilbert space average. 
To analyze this we consider the Hilbert space variance of, say, $\bra{\psi}\hat{S}\ket{\psi}$, i.e., $\Delta_{\text{H}}S= \Haver{\bra{\psi}\hat{S}\ket{\psi}^2}-\Haver{\bra{\psi}\hat{S}\ket{\psi}}^2$. 
If $\Delta_{\text{H}}S$ is small the above condition is satisfied. 
We would like to evaluate this for $S=:P_{ij,a}(t+\tau)-P_{ij,a}(t)$ under the restriction of given $P_{ij,a}(t)$. 
This, however, turns out to be mathematically rather involved and we have not managed to do so, yet. 
But, for the Hilbert space variance of any Hermitian operator $\hat{S}$ without any restriction one gets (cf.~\cite{Gemmer2004})
\begin{align}
  \label{eq:84}
  \Delta_{\text{H}}S
  = \frac{1}{N+1} 
    \Bigg[\frac{\trtxt{\hat{S}^2}}{N}-
    \Big(\frac{\trtxt{\hat{S}}}{N}\Big)^2\Bigg]\;,
\end{align} 
where the term in brackets obviously is the spectral variance of
$\hat{S}$ and $N$ denotes the dimension of the full system. 
At this point it simply appears plausible (which is of course far from being a proof) that the spectral variances of the above defined $S$ remain constant if one varies $N$, but keeps the rates $\gamma$ constant. 
Thus, for growing $N$ the replacement becomes more and more justified. 
Such a scenario is in accord with the general ideas of quantum thermodynamics as presented in \cite{Gemmer2004} and especially backed up by the numerical findings of \refsec{sec:4:subsec:2}. 

%
%
\section{Conclusion}
\label{sec:6}

Explicitly exploiting the Hilbert Space Average Method (HAM) we have in essence shown, that statistical relaxation may emerge directly from the Schr\"odinger equation. 
This requires the respective system being coupled in an adequate way to a suitable environment. 
This environment must feature many eigenstates. 
There is, however, no minimum particle number limit. 
Thus the thermodynamic limit appears to be essentially controlled by the number of environmental eigenstates involved in the dynamics rather than by the number of environmental particles. 
This relaxation behavior results even for correlated initial states, nevertheless, standard open system methods may fail to produce the correct result.

\begin{acknowledgement}
We are indebted to H.-P.~Breuer and G.~Mahler for interesting discussions on this subject. Financial Support by the Deutsche Forschungsgemeinschaft is gratefully acknowledged.
\end{acknowledgement}

\appendix

%
%
\section*{Appendix A: Time evolution}
\label{sec:7}

As already mentioned in \refsec{sec:2:subsec:1} in the interaction picture, $\hat{V}$ itself earns a time dependence
\begin{equation}
  \label{eq:22}
  \hat{V}(t) 
  = \E^{\I\hat{H}_{\text{loc}}t/\hbar}
    \,\hat{V}\,
    \E^{-\I\hat{H}_{\text{loc}}t/\hbar}\,.
\end{equation}
However, due to the organization of the interaction  (\ref{eq:5}) the total time dependence may be assigned only to the environment parts
\begin{equation}
  \label{eq:16}
  \hat{V}(t) = \sum_{ij}\hat{P}_{ij}\otimes\hat{C}_{ij}(t) 
\end{equation}
with
\begin{equation}
  \label{eq:16a} 
  \hat{C}_{ij}(t)
  =\E^{  \I\hat{H}_{\text{E}}t/\hbar} \hat{C}_{ij}
   \E^{- \I\hat{H}_{\text{E}}t/\hbar} \E^{\I(E_i-E_j)t/\hbar}\,.
\end{equation}
[We already mention this here for later reference, cf.~(\ref{eq:45})]. 
Assuming weak interactions (\ref{eq:18}) and short times $\tau$ the time evolution $\hat{D}(\tau, t)$ resulting from the Dyson series may be truncated at second order, i.e., 
\begin{equation}
  \label{eq:19}
  \ket{\psi(t+\tau)}
  \approx\underbrace{\Big[\hat{1}-\frac{\I}{\hbar}\hat{U}_1(\tau,t) 
  -\frac{1}{\hbar^2}\hat{U}_2(\tau,t)\Big]}_{\hat{D}_2(\tau, t)}
  \ket{\psi(t)}\;,
\end{equation}
with the two time evolution operators 
\begin{align}
  \label{eq:20}
  \hat{U}_1(\tau,t) &= \int_t^{\tau+t}\D\tau'\,\hat{V}(\tau'+t)\;, 
  \\
  \label{eq:21}
  \hat{U}_2(\tau,t) &= \int_t^{\tau+t}\D\tau'\int_t^{\tau'+t}\D\tau''
                    \,\hat{V}(\tau'+t) \,\hat{V}(\tau''+t)\;.
\end{align}
Note that the integration in (\ref{eq:21}) is time ordered, i.e., $\tau\geq\tau'\geq\tau''$.
Furthermore, the first order operator $\hat{U}_1(\tau)$ is Hermitian due to the Hermiticity of the interaction.

%
%
\section*{Appendix B: Correlation Functions}
\label{sec:8}

One has to analyze the traces on the right hand side (\ref{eq:36}).
Let us therefore abbreviate those term by $\bar{S}$. Using (\ref{eq:19}) we get\begin{align}
  \label{eq:37}
  &\bar{S}=\trtxt{\hat{D}_2^{\dagger}\proj{ij}{a}\hat{D}_2\proj{i'j'}{a'}}=
  \notag\\
  &\text{Tr}\Big\{\Big(\proj{ij}{a}
          +\frac{\I}{\hbar}\hat{U}_1\proj{ij}{a}
          -\frac{\I}{\hbar}\proj{ij}{a}\hat{U}_1
          +\frac{1}{\hbar^2}\hat{U}_1\proj{ij}{a}\hat{U}_1
  \notag\\
  &\phantom{\text{Tr}\Big\{}
          -\frac{1}{\hbar^2}\hat{U}_2^{\dagger}\proj{ij}{a}
          -\frac{1}{\hbar^2}\proj{ij}{a}\hat{U}_2\Big)
    \proj{i'j'}{a'}\Big\}\,,
\end{align}
where we used the Hermiticity of the operator $\hat{U}_1$.

To evaluate this complicated trace expression we will consider each order of time evolution operators in (\ref{eq:37}) separately, defining
\begin{equation}
  \label{eq:38}
  \bar{S}=S_0+S_1+S_2\;.
\end{equation}

Using (\ref{eq:11}) the zeroth order of (\ref{eq:37}) yields 
\begin{equation}
  \label{eq:40}
  S_0 
  = \trtxt{\proj{ij}{a}\proj{i'j'}{a'}}
  = \delta_{i'j}\delta_{j'i}\delta_{a'a}N_{a}\,.
\end{equation}

By a cyclic rotation within the trace the first order may be written as
\begin{align}
  \label{eq:41}
  S_1 
  &= \frac{\I}{\hbar}\trtxt{\hat{U}_1\proj{ij}{a}\proj{i'j'}{a'}-
                            \proj{i'j'}{a'}\proj{ij}{a}\hat{U}_1}
  \notag\\
  &=\frac{\I}{\hbar}\delta_{a'a}
     \big(\delta_{i'j} \trtxt{\hat{U}_1\proj{ij'}{a}}-
          \delta_{j'i} \trtxt{\hat{U}_1\proj{i'j}{a}}\big)\,,
\end{align}
where we used (\ref{eq:11}) again.
Concentrating on the first term, introducing the definition of the time evolution operator (\ref{eq:20}) and the interaction (\ref{eq:16}) one gets
\begin{align}
  \label{eq:42}
  &\trtxt{\hat{U}_1\proj{ij'}{a}}
  = \int_t^{\tau+t}\Dk\tau'\,\trtxt{\hat{V}(\tau'+t)\hat{P}_{ij'}\hat{\Pi}_a}
  \notag\\
  &= \int_t^{\tau+t}\Dk\tau'\,
     \sum_{kl}
     \trtxt{\hat{P}_{kl}\hat{C}_{kl}(\tau'+t)\hat{P}_{ij'}\hat{\Pi}_a}
  \notag\\
  &= \int_t^{\tau+t}\Dk\tau'\,
     \sum_{kl}
     \trtxt[\text{S}]{\hat{P}_{kl}\hat{P}_{ij'}}
     \trtxt[\text{E}]{\hat{C}_{kl}(\tau'+t)\hat{\Pi}_a}
  \notag\\
  &= \int_t^{\tau+t}\Dk\tau'\,
     \trtxt[\text{E}]{\hat{C}_{j'i}(\tau'+t)\hat{\Pi}_a}\,.
\end{align}
Due to the condition on the interaction (\ref{eq:8}) 
those terms are zero.
We find an analogous result for the second trace of (\ref{eq:41}) and thus we finally end up with 
\begin{equation}
  \label{eq:43}
  S_1=0\;.
\end{equation} 

For the second order terms of (\ref{eq:37}) we get 
\begin{align}
  \label{eq:44}
  S_2 
  =& \frac{1}{\hbar^2}\trtxt{\hat{U}_1\proj{ij}{a}\hat{U}_1\proj{i'j'}{a'}
  \notag\\
   &\phantom{\frac{1}{\hbar^2}\text{tr}}
   -\delta_{ji'}\delta_{a'a}\hat{U}_2^{\dagger}\proj{ij'}{a'}       
   -\delta_{j'i}\delta_{a'a}\hat{U}_2\proj{i'j}{a'}}\;.  
\end{align}
We concentrate first on the last term, plugging in the definition of $\hat{U}_2$ from (\ref{eq:21}) yields 
\begin{align}
  \label{eq:45}
  \trtxt{\hat{U}_2\proj{i'j}{a'}}
  =& \int_t^{\tau+t}\Dk\tau'\int_t^{\tau'+t}\Dk\tau''\times
  \notag\\
    &\times\trtxt{\hat{V}(\tau'+t)\hat{V}(\tau''+t)\proj{i'j}{a'}},
\end{align} 
exploiting (\ref{eq:16}) and performing the trace with respect to S we find
\begin{equation}
  \label{eq:45a}
  = \int_t^{\tau+t}\Dk\tau'\int_t^{\tau'+t}\Dk\tau''
     \sum_m
  \trtxt[\text{E}]{\hat{C}_{jm}(\tau'+t)\hat{C}_{mi'}(\tau''+t)\hat{\Pi}_{a'}}\,.
\end{equation}
Since the operators that generate the time-dependence of $\hat{V}(t)$ [cf.~(\ref{eq:16a})] commute with $\hat{\Pi}_{a'}$ and due to the invariance of the trace with respect to cyclic permutations of the traced operators, the above ``projected correlation functions'' only depend on the difference between the time arguments of the  $\hat{V}$'s. 
Since then the integrand no longer depends on $t$, the $t$ which appears in the integration boundaries may simply be set to zero. 
Hence one finds for the above expression
\begin{align}
  \label{eq:46}
  =\sum_m \int_0^{\tau}\D\tau'\int_0^{\tau'}\D\tau''
  \trtxt[\text{E}]{\hat{C}_{jm}(\tau'-\tau'')\hat{C}_{mi'}\hat{\Pi}_{a'}}\,.
\end{align}
As argued in the beginning the parts of the interaction are uncorrelated unless they are not adjoints of each other.
This means that the above traces can only be nonzero for the case $j=i'$.
Furthermore, one does the transformation $(\tau'-\tau'')\rightarrow\tau''$, thus, 
\begin{align}
  \label{eq:47}
  &= \sum_m \delta_{i'j}
     \int_0^{\tau}\D\tau'\int_0^{\tau'}\D\tau''
     \trtxt[\text{E}]{\hat{C}_{jm}(\tau'')\hat{C}_{mj}\hat{\Pi}_{a'}}\,.
\end{align}
Finally, plugging in the unit operator of the environment in terms of projection operators, one finds
\begin{align}
  \label{eq:48}
  &= \sum_m \delta_{i'j}
     \int_0^{\tau}\D\tau'\int_0^{\tau'}\D\tau''  
     \trtxt[\text{E}]{\hat{C}_{jm}(\tau'')
     \sum_b\hat{\Pi}_{b}\hat{C}_{mj}\hat{\Pi}_{a'}}
  \notag\\
  &= \sum_{mb} \delta_{i'j}
     \int_0^{\tau}\D\tau'\int_0^{\tau'}\D\tau''  
     \trtxt[\text{E}]{\hat{C}_{jm,a'b}(\tau'')
     \hat{C}_{mj,ba'}}\,.
\end{align}
Comparing this to (\ref{eq:39}) we end up with
\begin{equation}
  \label{eq:49}
  \trtxt{\hat{U}_2\proj{i'j}{a'}}
  = \sum_{mb} \delta_{i'j}\, f_{jm,a'b}(\tau)\;.
\end{equation}
Completely analogous we find for 
\begin{equation}
  \label{eq:50}
  \trtxt{\hat{U}_2^{\dagger}\proj{ij'}{a'}}
  = \sum_{mb} \delta_{j'i}\, f^*_{im,a'b}(\tau)\,.
\end{equation}

It remains the computation of the first term of (\ref{eq:44}).
Using the same argumentation as before (the fact that there are no correlations between different parts of the interaction as well as a cyclic rotation within the trace operation) we find for the trace
\begin{align}
  \label{eq:51}
  &\trtxt{\hat{U}_1\proj{ij}{a}\hat{U}_1\proj{i'j'}{a'}}
  = \int_t^{\tau+t}\Dk\tau'\int_0^{\tau+t}\Dk\tau''
    \delta_{ij}\,\delta_{i'j'}\times
  \notag\\
    &\quad\quad\times\,
    \trtxt[\text{E}]{\hat{C}_{ii'}(\tau''+t)\hat{\Pi}_{a'}\hat{C}_{i'i}(\tau'+t)\hat{\Pi}_{a}}
  \,.
\end{align}
By the same arguments which are given below  (\ref{eq:45}) this may be written independently of the absolute time $t$
\begin{align}
  \label{eq:52}
  &=\delta_{ij}\,\delta_{i'j'}\,
    \int_0^{\tau}\D\tau'\int_0^{\tau}\D\tau''   
    \trtxt[\text{E}]{\hat{C}_{ii'}(\tau''-\tau')\hat{\Pi}_{a'}\hat{C}_{i'i}\hat{\Pi}_{a}}
  \,.
\end{align}
The (non-time-ordered) integration of the above expression may be written in terms of a time-ordered integration by adding the time-reversed integrand
\begin{align}
  \label{eq:52a}
  &=\int_0^{\tau}\D\tau'\int_0^{\tau'}\D\tau''   
    \trtxt[\text{E}]{\hat{C}_{ii'}(\tau''-\tau')\hat{\Pi}_{a'}\hat{C}_{i'i}\hat{\Pi}_{a}}
  \notag\\
  &+\int_0^{\tau}\D\tau'\int_0^{\tau'}\D\tau'' \trtxt[\text{E}]{\hat{C}_{ii'}(\tau'-\tau'')\hat{\Pi}_{a'}\hat{C}_{i'i}\hat{\Pi}_{a}}
\,.
\end{align}
Shifting in the second term the time dependence to the other $\hat{C}$ operator and expressing everything within the trace by its adjoint yields
\begin{align}
  \label{eq:52b}
  &=\int_0^{\tau}\D\tau'\int_0^{\tau'}\D\tau''   
    \trtxt[\text{E}]{\hat{C}_{ii'}(\tau''-\tau')\hat{\Pi}_{a'}\hat{C}_{i'i}\hat{\Pi}_{a}}
  \notag\\
  &+\int_0^{\tau}\D\tau'\int_0^{\tau'}\D\tau'' 
  \trtxt[\text{E}]{(\hat{C}_{ii'}(\tau''-\tau')\hat{\Pi}_{a'}\hat{C}_{i'i}\hat{\Pi}_{a})^{\dagger}}
\,.
\end{align}
Since the trace of an adjoint operator is the complex conjugate of the original trace we may, after performing the same integral transformation described before (\ref{eq:47}), eventually write
\begin{equation}
  \label{eq:55}
  \trtxt{\hat{U}_1\proj{ij}{a}\hat{U}_1\proj{i'j'}{a'}}
  = \delta_{ij}\,\delta_{i'j'} 2\,\text{Re}\, f_{ii',aa'}(\tau)\,.
\end{equation}

Putting all the bits and peaces from (\ref{eq:40}), (\ref{eq:43}), (\ref{eq:44}), (\ref{eq:49}), (\ref{eq:50}) and (\ref{eq:55}) together, we eventually find
\begin{align}
  \label{eq:56}
  &\trtxt{\hat{D}_2^{\dagger}\proj{ij}{a}\hat{D}_2\proj{i'j'}{a'}}=\delta_{i'j}\delta_{j'i}\delta_{a'a}N_{a} \notag\\ 
&\Big(
  \delta_{ij}\,\delta_{i'j'}\, 2\mbox{Re} f_{ii',aa'}(\tau)
  \notag\\
  &- \sum_{mb} \delta_{j'i}\,\delta_{ji'}\,\delta_{a'a}\,
     \big[f^*_{im,a'b}(\tau)+f_{jm,a'b}(\tau)\big]\Big).
\end{align}



\end{document}